\documentclass[aps,pra,twocolumn,superscriptaddress,showpacs]{revtex4}
\usepackage{amsmath,amssymb,amsthm,graphicx}

\newcommand{\bra}[1]{\langle #1\vert}
\newcommand{\ket}[1]{\vert #1\rangle}
\newcommand{\braket}[2]{\langle #1\vert #2\rangle}
\newcommand{\ketbra}[1]{\vert #1\rangle\langle #1\vert}
\newcommand{\abs}[1]{\vert #1\vert}
\newcommand{\bigabs}[1]{\bigl\vert #1\bigr\vert}
\newcommand{\e}{\mathrm{e}}
\newcommand{\ir}{\mathrm{i}}
\newcommand{\one}{\openone}

\DeclareMathOperator{\tr}{Tr}

\newtheorem*{observation}{Observation}

\begin{document}

\title{Discrimination strategies for inequivalent classes of multipartite
entangled states}
\author{S\"onke Niekamp}
\affiliation{Institut f\"ur Quantenoptik und Quanteninformation,
\"Osterreichische Akademie der Wissenschaften,
Technikerstra\ss{}e 21a, A-6020 Innsbruck, Austria}
\author{Matthias Kleinmann}
\affiliation{Institut f\"ur Quantenoptik und Quanteninformation,
\"Osterreichische Akademie der Wissenschaften,
Technikerstra\ss{}e 21a, A-6020 Innsbruck, Austria}
\author{Otfried G\"uhne}
\affiliation{Institut f\"ur Quantenoptik und Quanteninformation,
\"Osterreichische Akademie der Wissenschaften,
Technikerstra\ss{}e 21a, A-6020 Innsbruck, Austria}
\affiliation{Institut f\"ur Theoretische Physik,
Universit\"at Innsbruck,
Technikerstra\ss{}e 25, A-6020 Innsbruck, Austria}
\date{\today}
\pacs{03.65.Wj, 03.67.Mn, 03.65.Ta}


\begin{abstract}
  How can one discriminate different inequivalent classes of multiparticle
  entanglement experimentally? We present an approach for the discrimination of
  an experimentally prepared state from the equivalence class of another state.
  We consider two possible measures for the discrimination strength of an
  observable. The first measure is based on the difference of expectation
  values, the second on the relative entropy of the probability distributions
  of the measurement outcomes. The interpretation of these measures and their
  usefulness for experiments with limited resources are discussed. In the case
  of graph states, the stabilizer formalism is employed to compute these
  quantities and to find sets of observables that result in the most decisive
  discrimination.
\end{abstract}

\maketitle

\section{Introduction}

With the rapid progress of quantum control, the experimental creation of a
variety of multiparticle entangled states has become
feasible~\cite{haeffner-2008-469,pan-2008}. When more than two particles are
entangled, it is well known that there are different and inequivalent
entanglement classes, but there are even different classification schemes: In a
first approach, one may consider two states $\ket{\psi}$ and $\ket{\phi}$ as
equivalent, if one can be converted into the other by changing the local bases
only. These operations are called local unitary (LU) operations and recently a
method to decide whether two states are LU equivalent or not has been
found~\cite{PhysRevLett.104.020504,Krau10}. Another classification is based on
the question regarding whether a single copy of $\ket{\psi}$ can be converted
into $\ket{\phi}$ by local operations and classical communication, even if this
conversion works only with a small probability~\cite{DuVC00}. These operations
are called stochastic local operations and classical communication (SLOCC).
Similar to LU equivalent states, states equivalent under SLOCC can often be
used for the same applications. As the number of SLOCC classes is infinite for
more than three qubits~\cite{VDMV02}, modified classification schemes have been
proposed~\cite{lamata-2006-74,BKPG09}.

On the theoretical side, it has been shown that different classes of entangled
states are suited for different applications. For example, cluster states are
useful for one-way quantum computation, whereas Greenberger-Horne-Zeilinger
(GHZ) states are not~\cite{nest-2006-97}. To the contrary, for sub shot-noise
interferometry, GHZ states are optimally suited, while cluster states are
useless for this task~\cite{HyGS09}. Consequently, it can be important to
discriminate experimentally between the different classes. 

For the experimental verification of entanglement, a number of tools
exist---the most prominent example are witness operators~\cite{GuTo09}. As
experiments no longer aim only at the creation of entanglement but also at
creating specific classes of entangled states, tools are needed for the
experimental discrimination of these classes. In the context of entanglement
detection, it is well known that a given Bell inequality or witness operator
detects only a part of all entangled states and fails to detect others. Thus
the violation (or nonviolation) of a Bell inequality can provide information
not only about the entanglement present in a state but also about its
type~\cite{PhysRevA.71.042325,KSWT05}.

Consequently, in Ref.~\cite{SKLW08} Bell operators  have been constructed and
experimentally implemented for discriminating different classes of entangled
states. For an experiment aiming at the creation of a particular state, a Bell
operator characteristic for this state was designed, that is, a Bell operator
that has the desired state as eigenstate with maximal eigenvalue. The maximal
expectation value of this Bell operator for various other classes of states
(defined as all LU equivalents or all SLOCC equivalents of some prominent
entangled state) was determined. Measuring the Bell operator then proved that
the prepared state was not in those classes with maximal expectation value
lower than the experimentally obtained value. In this approach, the
characteristic operator is far from unique. Neither is it necessary to use a
Bell operator, as has already been remarked in Ref.~\cite{SKLW08}. As entangled
states with increasingly large numbers of qubits are being prepared, analysis
tools that give strong results in spite of a limited number of measurement
events are needed. In the context of entanglement detection this problem has
recently received
attention~\cite{vDGG05,PhysRevLett.95.033601,PhysRevLett.104.210401}.

In this article we present an approach for the discrimination of an
experimentally prepared state from the class of all local unitaries of another
state. We define two measures for the discrimination strength of an observable.
The first measure is based of the difference of expectation values and
coincides with the one implicitly used in Ref.~\cite{SKLW08}. An interpretation
of this quantity as a noise tolerance is presented. The second measure is based
on the relative entropy, also called Kullback-Leibler divergence, which is a
well-established information-theoretic measure for the discrepancy of two
classical probability distributions~\cite{CoTh91,Vedr02}. This quantity is
directly related to the probability of another state to reproduce the observed
measurement outcomes in a given number of measurement runs. Our use of the
relative entropy is motivated by a work of van Dam~\emph{et al.}, where it was
used to assess the statistical strength of nonlocality proofs~\cite{vDGG05}. In
the case of graph states, the stabilizer formalism helps us to compute these
quantities and to find sets of observables that result in the strongest
discrimination. We would like to add that our approach is not directly related
to the task of state discrimination as it is often discussed in the
literature~\cite{chefles00}. In particular we do not assume the promise that
the state is either in the first or in the second family; such an assumption
cannot be justified in an experiment aiming for the verification of
entanglement properties.

This article is organized as follows: In Section~\ref{sec:approach} we
introduce the two measures for the discrimination strength. In
Sections~\ref{sec:four} and~\ref{sec:three} we calculate these quantities for
certain four- and three-qubit states and find optimal sets of observables for
the discrimination task. The performance of the measures for experimental data
and noisy states is investigated. In Section~\ref{sec:graph} a general result
for the discrimination of graph states is presented. Finally,
Section~\ref{sec:discussion} is devoted to a discussion of the results.

\section{The situation and the distance measures}
\label{sec:approach}

We consider the following situation: In an experiment aiming at the preparation
of a state $\varrho$ the experimenter wants to verify that the prepared state
is not in a certain class of undesired states, given by all local unitaries
(and maybe permutations of qubits) of a pure state $\ket{\phi}$. For that, he
can measure an observable $A$ (or several observables $A_i$) and one has to
define to what extent such a measurement can exclude the undesired states.

In this section we will define two quantities that measure how well an
observable $A$ discriminates a state $\varrho$ from all local unitaries of
another state $\ket{\phi}$. While we are restricting our attention to LU
classes in this article, it should be noted that the same quantities can also
be defined for SLOCC classes.

\subsection{A measure based on the fidelity}
\label{sec:fidelity}

In analogy to the approach taken in Ref.~\cite{SKLW08} we define the fidelity
based measure as
\begin{equation}
  \label{eq:diff}
  \mathcal{F}_A(\varrho\Vert\phi)
  =\min_{U\in\text{LU}}\bigabs{\tr(A\varrho)-\bra{\phi}U^\dagger AU\ket{\phi}}.
\end{equation}
Here the minimization is over all local unitaries $U$; later we will consider
also the minimization over all permutations of qubits.

Adding or subtracting a multiple of the identity matrix to $A$ does not change
the value of $\mathcal{F}$ in Eq.~\eqref{eq:diff}. So without loss of
generality we can assume $\tr(A)=0$. Also without loss of generality we assume
$\tr(A\varrho)\ge 0$. 

In the following we will always assume that $\tr(A\varrho)\ge
\max_\text{LU}\bra{\phi}A\ket{\phi}$. This is no restriction for our purposes
due to the following reasoning: Let us assume that for the  pure $n$-qubit
state $\ket{\phi}$ there exist $2^n$ local unitaries $U_i$ such that
$\{U_i\ket{\phi}\}$ forms an orthonormal basis. States with this property are
called locally encodeable~\cite{TaMM07}. It has been conjectured that all pure
states are locally encodeable, and the conjecture has been proven for a variety
of states, including all stabilizer states and the W state~\cite{TaMM07}. Then
we can write $\one=\sum_iU_i\ketbra{\phi}U_i^\dagger$ and since $A$ is
traceless we have $\tr(A\sum_iU_i\ketbra{\phi}U_i^\dagger)=0$. So, if
$\bra{\phi}A\ket{\phi}<0$ there exists a local unitary $U$ such that
$\bra{\phi}U^\dagger AU\ket{\phi}>0$ and vice versa. If
$\max_\text{LU}\bra{\phi}A\ket{\phi}\ge\tr(A\varrho)$, by local encodeability
there exists a local unitary $U$ such that $\tr(A\varrho)\ge
0\ge\bra{\phi}U^\dagger AU\ket{\phi}$. By continuity there exists another local
unitary such that $\tr(A\varrho)=\bra{\phi}U^\dagger AU\ket{\phi}$, i.\,e.,
$\mathcal{F}_A(\varrho\Vert\phi)=0$ and the observable $A$ is not suitable for a
discrimination procedure based on $\mathcal{F}$. In this article, we shall be
concerned with graph states and sometimes with the W state, so local
encodeability is proven for our purposes and we have 
\begin{equation}
  \label{eq:diff2}
  \mathcal{F}_A(\varrho\Vert\phi) 
  =\tr(A\varrho)-\max_\text{LU}\bra{\phi}A\ket{\phi}.
\end{equation}

This quantity, however, is not invariant under rescaling of $A$. To be able to
compare different observables, we have to agree on a normalization. We choose
$\tr(A\varrho)=1$, which is the same normalization as in Ref.~\cite{SKLW08},
and obtain
\begin{equation}
  \mathcal{F}_A(\varrho\Vert\phi)=1-\max_\text{LU}\bra{\phi}A\ket{\phi}.
\end{equation}
This is the first quantity that will serve us as a measure for the strength
with which $A$ discriminates $\varrho$ from all local unitaries of
$\ket{\phi}$.

For more than one observable we define
\begin{equation}
  \mathcal{F}_{A_1,\ldots,A_k}(\varrho\Vert\phi)
  =\mathcal{F}_{\frac{1}{k}\sum_{i=1}^kA_i}(\varrho\Vert\phi).
\end{equation}
In the remainder of the article we will discuss how to find optimal families of
observables $A_1,\ldots,A_k$ for particular states $\varrho$ and $\ket{\phi}$.

Our definition has a direct physical interpretation in terms of a noise
tolerance. For that, we exploit the similarity of our problem to the task of
entanglement detection by virtue of witness operators and consider the
robustness of $\mathcal{F}$ in Eq.~\eqref{eq:diff2} against white noise: Let
\begin{equation}
  \varrho_\text{wn}(p)=(1-p)\frac{\one}{d}+p\varrho
\end{equation}
be the state $\varrho$ affected by white noise. The maximal noise level $(1-p)$
such that
\begin{equation}
  \tr[A\varrho_\text{wn}(p)]-\max_\text{LU}\bra{\phi}A\ket{\phi}\ge 0
\end{equation}
is given by [using $\tr(A)=0$]
\begin{equation}
  1-p=1-\max_\text{LU}\bra{\phi}A\ket{\phi}=\mathcal{F}_A(\varrho\Vert\phi).
\end{equation}
We will discuss the interpretation of $\mathcal{F}$ as a noise tolerance in
more detail in Section~\ref{sec:examples}.

\subsection{A measure based on the relative entropy}
\label{sec:relent}

{From} a statistical point of view, the task of discriminating a state
$\varrho$ and a state $\sigma=\ketbra{\phi}$ by virtue of an observable $A$ is
the task of discriminating the corresponding probability distributions for the
measurement outcomes of $A$.

The relative entropy or Kullback-Leibler divergence is a well-established
information-theoretic measure for the discrepancy between two classical
probability distributions~\cite{Vedr02,CoTh91}. The relative entropy of the
probability distributions $P=\{p_1,\ldots,p_m\}$ and $Q=\{q_1,\ldots,q_m\}$ is
defined as
\begin{equation}
  D(P\Vert Q)=\sum_{i=1}^mp_i\log\bigl(\frac{p_i}{q_i}\bigr).
\end{equation}
We will always use the logarithm to the base of two, $\log=\log_2$, and define
$0\log(0)=0$. Note that since we are dealing with the discrimination of
classical probability distributions, we are not using the quantum (or von
Neumann) relative entropy~\cite{Vedr02},
$\tr[\varrho(\log\varrho-\log\sigma)]$. The relative entropy satisfies $0\le
D(P\Vert Q)\le\infty$ with $D(P\Vert Q)=0$ if and only if $P=Q$. But although
the relative entropy behaves in some sense like a distance between probability
distributions, it is not a metric because it is not symmetric. The most
important properties of the relative entropy are summarized in
Appendix~\ref{app:relent}.

Concerning the interpretation, the relative entropy $D(P\Vert Q)$ can be used
to answer the question: How strongly does a sample (of a fixed length) from the
distribution $P$ on average indicate that it was indeed drawn from $P$ rather
than from $Q$? This statement can be made precise with the theory of
statistical hypothesis testing~\cite{vDGG05}.

For the simplest case, note that $D(P\Vert Q)$ is infinity if and only if
$q_i=0$, but $p_i>0$ for some $i$, that is, if an event is impossible according
to $Q$, but occurs with a nonvanishing probability according to $P$. This means
that on observing this event one immediately knows that the sample was not
drawn from $Q$.

More generally, suppose that a sample of length $N$ has been drawn from $Q$. We
consider the empirical probability distribution $P$ defined by the observed
frequencies. Then the probability $Q^N[T(P)]$ of drawing a sample from $Q$ with
the same frequencies [i.\,e., within the type class  $T(P)$] decays
exponentially for large $N$ \cite[Thm.~12.1.4]{CoTh91},
\begin{equation}
  \label{eq:typeprob}
  Q^N[T(P)]\sim 2^{-ND(P\Vert Q)}.
\end{equation}
Consequently, if one observes a probability distribution $P$ yielding a large
value for the relative entropy $D(P\Vert Q)$, the assumption that it was rather
drawn from the probability distribution $Q$ is very questionable (see also
below).

Let us now return to our original problem: For an experiment aiming at the
preparation of the state $\varrho$, we define a measure for how well the
observable $A$ can exclude the state $\sigma$ as the relative entropy of the
corresponding measurement outcomes for $A$
\begin{equation}
  \label{eq:protod}
  D_A(\varrho\Vert\sigma)
  =\sum_{i=1}^m\tr(\varrho\Pi_i)
  \log\bigl(\frac{\tr(\varrho\Pi_i)}{\tr(\sigma\Pi_i)}\bigr),
\end{equation}
where $A=\sum_{i=1}^ma_i\Pi_i$ is the spectral decomposition of $A$. {From} the
above discussion, this is a measure for how strongly the measurement results of
the observable $A$ on the state $\varrho$ on average show that they are due to
the state $\varrho$ rather than the state $\sigma$. (Note, that in this
interpretation we assume that the experimental precision in implementing the
observable $A$ outperforms the precision that can be achieved for the
preparation of the state $\varrho$.)

Let us discuss the interpretation of this quantity. Suppose that the
measurement has been performed, resulting in an observed probability
distribution $\tilde{P}=\{\tilde{p}_1,\ldots,\tilde{p}_m\}$ of the outcomes
$a_1,\ldots,a_m$, and let $\tilde{N}$ be the number of measurement runs. Then,
by Eq.~\eqref{eq:typeprob}, the probability that a measurement on the state
$\sigma$, after $\tilde{N}$ measurement runs, results in the same frequencies
is given by
\begin{equation}
  \label{eq:stateprob}
  Q^{\tilde{N}}[T(\tilde{P})]\sim 2^{-\tilde{N}D(\tilde{P}\Vert Q)},
\end{equation}
where $Q=\{\tr(\sigma\Pi_1),\ldots,\tr(\sigma\Pi_m)\}$. If the experimentally
prepared state is close enough to the intended state $\varrho$, the relative
entropy $D(\tilde{P}\Vert Q)$ will attain a large value only if this is already
the case for $D_A(\varrho\Vert\sigma)$.

For comparison, when tossing a fair coin $N$ times, the probability of the
outcome always being ``tails'' is
\begin{equation}
  2^{-ND(\{1,0\}\Vert\{\frac{1}{2},\frac{1}{2}\})}=2^{-N}
\end{equation}
since Eq.~\eqref{eq:typeprob} is exact in this example. Thus, the probability
Eq.~\eqref{eq:stateprob} of obtaining the frequencies $\tilde{P}$ after
measuring $\tilde{N}$ times the state $\sigma$ is equal to the probability of
always obtaining ``tails'' in $N=\tilde{N}D(\tilde{P}\Vert Q)$ tosses of a fair
coin~\cite{vDGG05}. In other words, the likelihood after $\tilde{N}$
measurement runs that the prepared state is $\sigma$ is the same as that of a
coin to be fair after $N=\tilde{N}D(\tilde{P}\Vert Q)$ tosses resulting in
``tails''. This gives our results for the measure $D$ in Eq.~\eqref{eq:protod}
a quantitative interpretation.

When measuring several observables $A_1,\ldots,A_k$ independently of each
other, the relative entropy of the joint probability distributions is given by
the sum of the relative entropies for the individual observables (cf.
Property~\ref{it:additivity} in Appendix~\ref{app:relent}). However, we
renormalize the relative entropy in this case and define
\begin{equation}
  \label{eq:several}
  D_{A_1,\ldots,A_k}(\varrho\Vert\sigma)
  =\frac{1}{k}\sum_{i=1}^kD_{A_i}(\varrho\Vert\sigma),
\end{equation}
where the prefactor $1/k$ corresponds to keeping the overall number of
measurement runs constant, independent of the number of observables, i.\,e.,
each observable $A_i$ will be measured in $\tilde{N}/k$ runs. We choose this
definition because in experiments the rate at which entangled states are being
created is typically low, so the number of measurement runs is a scarce
resource. 

Finally, we consider the minimum of $D$ over all local unitaries of $\sigma$,
\begin{equation}
  \label{eq:d}
  \mathcal{D}_{A_1,\ldots,A_k}(\varrho\Vert\sigma)
  =\min_{U\in\text{LU}}D_{A_1,\ldots,A_k}(\varrho\Vert U\sigma U^\dagger).
\end{equation}
In the following we will discuss how to find families of observables $A_i$
which maximize this quantity.

\section{Discriminating four-qubit states}
\label{sec:four}

As our first example, we will calculate the quantities $\mathcal{F}$ and
$\mathcal{D}$ for the discrimination of the four-qubit GHZ state from the
four-qubit linear cluster state and vice versa.

The four-qubit GHZ state is given by 
\begin{equation}
  \ket{\text{GHZ}_4}=\frac{1}{\sqrt{2}}\bigl(\ket{0000}+\ket{1111}\bigr).
\end{equation}
Alternatively, it can be described by its stabilizing operators: The GHZ state
is the unique common eigenstate with eigenvalue $+1$ of the 16 operators
\begin{multline}
  \label{eq:ghzstab}
  S_{\text{GHZ}_4}
  =\{\one\one\one\one,\ \one\one ZZ \text{ and perm.},\ ZZZZ,\\
  XXXX,\ -XXYY\text{ and perm.},\ YYYY\}.
\end{multline}
Here and in the following $X$, $Y$, $Z$, and $\one$ denote the Pauli matrices
and the identity, tensor product signs have been omitted and ``perm'' denotes
all possible permutations of the qubits which give different terms. The set of
all stabilizing operators forms a commutative group, and the GHZ state is an
example of a graph state~\cite{HDER06}. We will explain this in more detail in
Section~\ref{sec:graph} below. The sum of all stabilizing operators gives the
projector onto the state
\begin{equation}
  \label{eq:gsproj}
  \ketbra{\text{GHZ}_4}=\frac{1}{16}\sum_{S\in S_{\text{GHZ}_4}}S,
\end{equation}
this property is shared by any graph state. The stabilizing operators thus
contain a description of the correlations present in the state.

The linear cluster state  given by \begin{equation}
  \ket{\text{C}_4}
  =\frac{1}{2}\bigl(\ket{0000}+\ket{0011}+\ket{1100}-\ket{1111}\bigr)
\end{equation}
is also a graph state: Its stabilizer group is~\cite{HDER06}
\begin{equation}
  \label{eq:clstab}
  \begin{split}
    S_{\text{C}_4}
    =\{&\one\one\one\one,\ \one\one ZZ,\ ZZ\one\one,\ ZZZZ,\\
    & XYXY,\ XYYX,\ YXXY,\ YXYX,\\
    &\one ZXX,\ Z\one XX,\ XX\one Z,\ XXZ\one,\\
    &{-\one ZYY},\ -Z\one YY,\ -YY\one Z,\ -YYZ\one\},
  \end{split}
\end{equation}
and the analogous relation to Eq.~\eqref{eq:gsproj} holds.

The stabilizing operators of a graph state $\ket{\psi}$ provide a natural
choice of observables for the discrimination of $\varrho=\ketbra{\psi}$ from
other states. In the language of Ref.~\cite{SKLW08}, they are characteristic
operators for the graph state. In the following we will restrict our analysis
to these observables.

\subsection{Discriminating the GHZ state from the cluster state}

We first consider the discrimination of the GHZ state from all LU equivalents
of the cluster state, using \emph{all} stabilizing operators of the former,
excluding only the identity, as it is useless for any discrimination task. We
introduce the notation $S^*=S\setminus\{\one\}$ for any stabilizer group $S$
minus the identity. Later we will discuss which \emph{subset} of the stabilizer
group gives the strongest discrimination. We start with the calculation of the
quantity $\mathcal{D}$ in Eq.~\eqref{eq:d}, which is based on the relative
entropy.

It is useful to think of the cluster state as the sum of its stabilizing
operators $\ketbra{\text{C}_4}=\frac{1}{16}\sum_{T\in S_{\text{C}_4}}T$. For
any GHZ stabilizing operator $S\in S^*_{\text{GHZ}_4}$, the term $D_S$ is a
function of the overlap of $S$ with the stabilizing operators of the cluster
state
\begin{equation}
  D_S(\text{GHZ}_4\Vert\text{C}_4)
  =-\log\Bigl\{\frac{1}{2}\Bigl[\frac{1}{16}\sum_{T\in S_\text{C}}\tr(ST)
  +1\Bigr]\Bigr\}.
\end{equation}
If we do not consider local unitaries, $\tr(ST)$ is zero unless $S=T$. For the
minimization over local unitaries in Eq.~\eqref{eq:d} we classify stabilizing
operators by the qubits on which they act nontrivially. For any GHZ stabilizing
operator $S$, only those stabilizing operators of $\ket{\text{C}_4}$ which act
nontrivially on the exactly the same qubits as $S$ can have a nonvanishing
overlap with $S$. This still holds if arbitrary local unitary operations are
applied to $\ket{\text{C}_4}$. We can thus identify those stabilizing operators
of $\ket{\text{C}_4}$ that can contribute to $D_S$.

If no local unitary is applied, the GHZ stabilizing operators $\one\one ZZ$,
$ZZ\one\one$, and $ZZZZ$ have maximal overlap with cluster stabilizing
operators and thus give the minimal relative entropy of zero, while $\one
Z\one Z$, $\one ZZ\one$, $Z\one\one Z$, and $Z\one Z\one$ each have zero
overlap and thus give relative entropy of 1. A minimization over local
unitaries cannot improve this result, as the cluster state has no stabilizing
operators acting nontrivially on the same qubits.

All of the remaining stabilizing operators of $\ket{\text{GHZ}_4}$
\begin{multline}
  \Sigma =\{XXXX,\ -XXYY,\ -YYXX,\ YYYY,\\
  {-XYXY},\ -XYYX,\ -YXXY,\ -YXYX\}
\end{multline}
act on all four qubits (such stabilizing operators describing four-point
correlations we call four-point stabilizing operators for short). We note that
both these and the four-point stabilizing operators of $\ket{\text{C}_4}$
except $ZZZZ$ are products of local operators $X$ and $Y$. It is therefore
reasonable to assume that for the minimization of $D_\Sigma$ it suffices to
consider rotations about the $z$ axes. The rotated cluster state is
\begin{multline}
  \ket{\text{C}_4(\gamma,\delta)}
  =\frac{1}{2}\bigl(\ket{0000}+\e^{-\ir\delta}\ket{0011}\\
  +\e^{-\ir\gamma}\ket{1100}-\e^{-\ir(\gamma+\delta)}\ket{1111}\bigr),
\end{multline}
where $\gamma=\varphi_1+\varphi_2$, $\delta=\varphi_3+\varphi_4$, and the
$\varphi_i$ are the rotation angles about the local $z$ axes, and we obtain
$D_\Sigma(\text{GHZ}_4\Vert\text{C}_4(\gamma,\delta))
=-\frac{1}{2}\{\log[\frac{1}{2}(1+\sin(\gamma)\sin(\delta))]
+\log[\tfrac{1}{2}(1-\cos(\gamma)\cos(\delta))]\}$. The minimum of this
expression is $-\log(3/4)$. In conclusion, we have found that
\begin{equation}
  \label{eq:ghzcd}
  \mathcal{D}_{S^*_{\text{GHZ}_4}}(\text{GHZ}_4\Vert\text{C}_4)
  =\frac{1}{15}\bigl(4-8\log\frac{3}{4}\bigr)\approx 0.4880.
\end{equation}
Since our analytic optimization required an assumption we would like to add
that this result is also obtained via numerical minimization over all local
unitaries. The 15 GHZ stabilizing operators do not contribute equally to
$\mathcal{D}$, rather, $D=0$ for  $\one\one ZZ$, $ZZ\one\one$, and $ZZZZ$;
$D=1$ for $\one Z\one Z$, $\one ZZ\one$, $Z\one\one Z$, and $Z\one Z\one $; and
$D=-\log(3/4)\approx 0.4150$ for all others.

Let us now turn to the fidelity-based measure $\mathcal{F}$ for the same
observables and states. If we use all stabilizing operators of $\ket{\psi}$,
excluding again only the identity, $\mathcal{F}(\psi\Vert\phi)$ is a function
of the fidelity
\begin{equation}
  \label{eq:fid}
  \mathcal{F}_{S^*_\psi}(\psi\Vert\phi)
  =\frac{2^n}{2^n-1}\Bigl(1-\max_\text{LU}\abs{\braket{\psi}{\phi}}^2\Bigr),
\end{equation}
where $n$ is the number of qubits.

For our example, we note that for an arbitrary local unitary $U$ we have
$\abs{\bra{\text{GHZ}_4}U\ket{\text{C}_4}}^2 \le 1/2$ and this bound can be
reached. This follows from the known fact that the maximal overlap of the
cluster state with any product state, and thus with $\ket{0000}$ and
$\ket{1111}$, is given  by $1/4$~\cite{markham-2007-9} and one can easily find
a local unitary with $\abs{\bra{\text{GHZ}_4}U\ket{\text{C}_4}}^2=1/2.$ So we
have
\begin{equation}
  \label{eq:ghzcf}
  \mathcal{F}_{S^*_{\text{GHZ}_4}}(\text{GHZ}_4\Vert\text{C}_4) =\frac{8}{15}
\end{equation}
as the fidelity-based measure for the discrimination.

Let us now discuss subsets of the stabilizer group as observables for the
discrimination. In our previous analysis, it turned out that  not all
stabilizing operators contribute equally to the discrimination, in fact, some of
them do not contribute at all. We therefore ask for families of stabilizing
operators of $\ket{\text{GHZ}_4}$ that discriminate $\ket{\text{GHZ}_4}$ from
the local unitaries of $\ket{\text{C}_4}$ most strongly, that is, families for
which $\mathcal{F}$ (or $\mathcal{D}$) is maximal. 

{From} the previous discussion, candidates are
\begin{equation}
  \label{eq:opt}
  \one Z\one Z,\ \one ZZ\one,\ Z\one\one Z,\ Z \one Z\one,
\end{equation}
since any of them gives $\mathcal{F}=\mathcal{D}=1$. But if we want to exclude
not only all LU equivalents, but also all permutations of qubits of the cluster
state, we still have to minimize $\mathcal{F}$ and $\mathcal{D}$ over all
permutations, because the set of observables is no longer necessarily
permutation invariant. There are three distinct permutations of the cluster
state, namely
$\ket{\text{C}^1}=\ket{\text{C}_4}
=\tfrac{1}{2}(\ket{0000}+\ket{0011}+\ket{1100}-\ket{1111})$,
$\ket{\text{C}^2}=\tfrac{1}{2}(\ket{0000}+\ket{0110}+\ket{1001}-\ket{1111})$,
and
$\ket{\text{C}^3}=\tfrac{1}{2}(\ket{0000}+\ket{0101}+\ket{1010}-\ket{1111})$.
Table~\ref{tab:optimal} shows the GHZ stabilizing operators from
Eq.~\eqref{eq:opt} along with all stabilizing operators of the permutations of
$\ket{\text{C}_4}$ that act on the same qubits. We see that any single one of
these six stabilizing operators gives a relative entropy of zero, if the
entropy is minimized over all permutations. Any pair of stabilizing operators
gives an entropy of either zero or $1/2$. The three-element family $\{\one\one ZZ,
\one Z\one Z,\one ZZ\one\}$ gives $2/3$, in total there are eight such families
giving the same value.

\begin{table}
  \centering
  \begin{tabular}{c|ccc}
    $\ket{\text{GHZ}_4}$ & $\ket{\text{C}^1}$ &
    $\ket{\text{C}^2}$ & $\ket{\text{C}^3}$\\
    \hline
    $\one\one ZZ$ & $\one\one ZZ$ & & \\ 
    $\one Z\one Z$ & & & $\one Z\one Z$\\ 
    $\one ZZ\one $ & & $\one ZZ\one$ & \\	
    $Z\one\one Z$ & & $Z\one\one Z$ & \\
    $Z\one Z\one $ & & & $Z\one Z\one$\\
    $ZZ\one\one $ & $ZZ\one\one$ & & \\
  \end{tabular}
  \caption{\label{tab:optimal} Stabilizing operators of the GHZ state and
  stabilizing operators of the three permutations of the cluster state acting
  on the same qubits (see text for further details).}
\end{table}

It is easy to see that these families of stabilizing operators are optimal: It
is clear that they are optimal among all subsets of the six stabilizing operators
in the table. Furthermore, we recall that we found a local unitary
transformation such that all remaining stabilizing operators contribute either
$0$ or $-\log(3/4)$ to the entropy. Because of the permutation invariance of
the set of these remaining stabilizing operators, this holds for all
permutations of $\ket{\text{C}_4}$. As $-\log(3/4)<2/3$, adding some of the
remaining observables cannot improve the discrimination. This shows the
optimality of our three-element families. These families are also optimal when
using $\mathcal{F}$ instead of $\mathcal{D}$.

We summarize the main results of this subsection in the following observation:

\begin{observation}
  For the discrimination of the GHZ state from all local unitaries and
  permutations of qubits of the cluster state, using all GHZ stabilizing
  operators except the identity, the measures $\mathcal{D}$ and $\mathcal{F}$
  are given by Eqs.~\eqref{eq:ghzcd} and~\eqref{eq:ghzcf}. When considering
  subsets of the stabilizer group, $\{\one\one ZZ,\ \one Z\one Z,\ \one
  ZZ\one\}$ is an example of an optimal family of observables, giving
  $\mathcal{F}=\mathcal{D}=2/3$.
\end{observation}

Finally, let us add that until now we assumed that all observables are measured
independently. However, as the observables in Eq.~\eqref{eq:opt}, from which we
constructed the optimal families, have a common eigenbasis of product states
(the computational basis), they can also be measured jointly in one experiment
with more than two outcomes. In this case, the relative entropy is no longer
given by Eq.~\eqref{eq:several}. When measuring the computational basis in one
experiment with 16 outcomes, we obtain
\begin{equation}
  \label{eq:bases}
  \begin{split}
    \mathcal{D}
    &=\min_{U\in\text{LU}}\sum_{i=1}^{16}
    \abs{\braket{e_i}{\text{GHZ}_4}}^2
    \log\Bigl(\frac{\abs{\braket{e_i}{\text{GHZ}_4}}^2}
    {\abs{\bra{e_i}U\ket{\text{C}_4}}^2}\Bigr)\\
    &=1.
  \end{split}
\end{equation}
Consequently, considering measurements with more outcomes can give a stronger
discrimination. This is a consequence of a general feature of the relative
entropy: For each of the observables in Eq.~\eqref{eq:opt}, the probability
distribution for the measurement outcomes is obtained from the one for the
measurement of the computational basis by considering several events as one (in
other words, by ``forgetting'' information). The relative entropy satisfies a
grouping rule similar to the Shannon entropy (Property~\ref{it:grouping} in
Appendix~\ref{app:relent}), which implies that this process can only decrease
the relative entropy. 

\subsection{Discriminating the cluster state from the GHZ state}

Let us now consider the reverse discrimination
$\mathcal{D}_{S^*_{\text{C}_4}}(\text{C}_4\Vert\text{GHZ}_4)$. This turns out
to be relatively simple.

First, note that the eight three-point stabilizing operators of
$\ket{\text{C}_4}$ will for any local unitary operation have zero overlap with
$\ket{\text{GHZ}_4}$ as the GHZ state has no three-point stabilizing operators.
For the remaining eight stabilizing operators, however, the overlap with the
GHZ state can brought to $1$ by an appropriate rotation, as one can directly
check. Thus
\begin{equation}
  \mathcal{D}_{S^*_{\text{C}_4}}(\text{C}_4\Vert\text{GHZ}_4)=\frac{8}{15}.
\end{equation}
As a function of the fidelity, $\mathcal{F}$ is the same as for the reverse
discrimination
\begin{equation}
  \mathcal{F}_{S^*_{\text{C}_4}}(\text{C}_4\Vert\text{GHZ}_4)
  =\frac{8}{15}.
\end{equation}

Considering the optimal subsets of the stabilizer group, it is clear that any
set of three-point stabilizing operators of $\ket{\text{C}_4}$ is an optimal
family of observables, resulting in $\mathcal{D}=\mathcal{F}=1$. Note that the
GHZ state is permutation invariant, so the optimization over permutations does
not play a role.

\subsection{Application to a four-photon experiment}
\label{sec:examples}

To study the noise tolerance of the quantities $\mathcal{F}$ and $\mathcal{D}$
and their performance for experimental data, we use the measurement results for
the stabilizer correlations of the cluster state obtained by Kiesel~\emph{et
al.} in a photonic experiment~\cite{KSWT05}. When using all cluster stabilizing
operators for the discrimination (excluding the identity), these data give
\begin{align}
  \label{eq:exp1}
  \mathcal{F}_{S^*_{\text{C}_4}}(\varrho_\text{exp}\Vert\text{GHZ}_4)
  &=0.257\pm 0.014,\\
  \label{eq:exp2}
  \mathcal{D}_{S^*_{\text{C}_4}}(\varrho_\text{exp}\Vert\text{GHZ}_4)
  &=0.189\pm 0.012.
\end{align}
When using only the three-point stabilizing operators, which form an optimal family
$Q$, we get
\begin{align}
  \label{eq:exp3}
  \mathcal{F}_Q(\varrho_\text{exp}\Vert\text{GHZ}_4)&=0.668\pm 0.019,\\
  \label{eq:exp4}
  \mathcal{D}_Q(\varrho_\text{exp}\Vert\text{GHZ}_4)&=0.353\pm 0.021.
\end{align}
Note that in all cases the observables are normalized with respect to the
perfect cluster state as $\bra{\text{C}_4}A_i\ket{\text{C}_4}=1$, while for the
experimental data we have $\tr(\varrho_\text{exp}A_i)<1$. Also, it should be
noted that the subsets of observables we use were chosen to be optimal for the
perfect cluster state but not necessarily for the experimental one. This,
however, is similar to the implementation of entanglement witnesses in
experiments: There, one typically considers some optimal witness for some pure
state that one aims to prepare and applies it to the experimental data in
order to obtain a significant entanglement test~\cite{GuTo09}.

\begin{figure}
  \centering
  \includegraphics{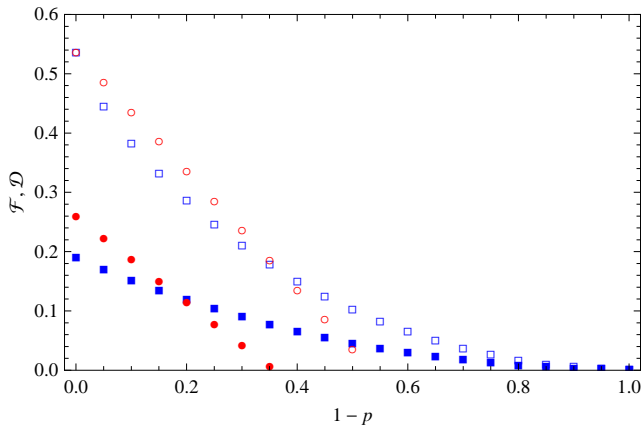}
  \caption{\label{fig:all} Discriminating the four-qubit linear cluster state
  with noise from all local unitaries of the GHZ state, using all stabilizing
  operators of the former. Shown are $\mathcal{F}$ (red circles) and
  $\mathcal{D}$ (blue squares) versus the level of white noise $(1-p)$ for the
  perfect (empty symbols) and the experimental (filled symbols) state.}
\end{figure}

\begin{figure}
  \centering
  \includegraphics{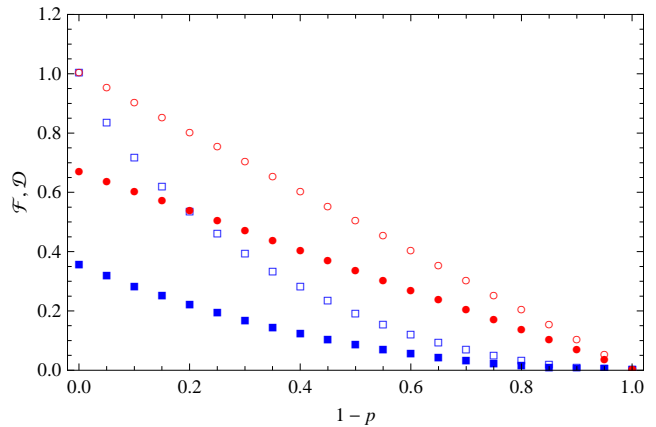}
  \caption{\label{fig:threepoint}  Discriminating the four-qubit linear cluster
  state with noise from all local unitaries of the GHZ state, using all
  three-point stabilizing operators of the former. Symbols as in
  Fig.~\ref{fig:all}.}
\end{figure}

To investigate the power of our discrimination methods, we calculate
$\mathcal{F}$ and $\mathcal{D}$ for both the perfect and the experimental
cluster state under the influence of white noise. Figures~\ref{fig:all}
and~\ref{fig:threepoint} show $\mathcal{F}$ and $\mathcal{D}$ as functions of
the noise level. We make a number of observations:

For the perfect cluster state with additional white noise, the quantity
$\mathcal{F}$ decreases with increasing noise level until it reaches zero at
the noise level $(1-p)=\mathcal{F}(\text{C}_4\Vert\text{GHZ}_4)$. This
interpretation of $\mathcal{F}$ as a noise tolerance was already mentioned in
Section~\ref{sec:fidelity}. Note, however, that the in the case of the
experimental state the noise tolerance is no longer given by
$\mathcal{F}(\varrho_\text{exp}\Vert\text{GHZ}_4)$ but is larger due to
$\tr(\varrho_\text{exp}A_i)<1$.

For the same observables, the maximal noise level at which $\mathcal{D}>0$ is
at least as high as the maximal noise level at which $\mathcal{F}>0$. This is a
general feature: As a consequence of the positive definiteness of the relative
entropy, $D$ is nonzero whenever $F$ is.  In this particular example,
$\mathcal{D}$ is nonzero for noise levels arbitrarily close to 1, though this
is not a general feature.

For the three-point stabilizing operators of $\ket{\text{C}_4}$, also the noise
tolerance of $\mathcal{F}$ is 1 (Fig.~\ref{fig:threepoint}). It is
instructive to compare this to the case of witness operators: The set of
separable states contains a ball around the completely mixed
state~\cite{PhysRevLett.83.1054}, which implies that for any witness $W$ and
entangled state $\varrho$ detected by $W$ the noise tolerance is strictly less
than 1. In our case the situation is different: The reason for the noise
tolerance of one is that no local unitaries of the GHZ state have any
three-point correlations. This implies that the set of states LU equivalent to
$\ket{\text{GHZ}_4}$ does not contain a ball around $\one/d$. For a fair
comparison, one may therefore consider the three-point correlations in
experiments \emph{aiming} at the generation of GHZ states (for instance, in
Ref.~\cite{WSKP08} they were maximally 0.097) and ask whether the measured
three-point stabilizer correlations in a cluster state experiment significantly exceed
these values.

Comparing the figures shows another difference between the measures
$\mathcal{F}$ and $\mathcal{D}$: For the measure $\mathcal{F}$, the noise
tolerance is higher in the case of the three-point stabilizing operators
(Fig.~\ref{fig:threepoint}) than in the case of all stabilizing operators
(Fig.~\ref{fig:all}). This is remarkable because the former set of observables
is contained in the latter. In other words, adding an observable can reduce the
noise tolerance of $\mathcal{F}$. From the definition of the quantity
$\mathcal{D}$ it is clear that its noise tolerance of
$\mathcal{D}_{A_1,\ldots,A_k}$ is lower bounded by the noise tolerance of
$\mathcal{D}$ for any subset of $A_1,\ldots,A_k$. In Ref.~\cite{BBLP08}, a
quantity similar to $\mathcal{F}$ was constructed from the correlations of an
entangled state and used for entanglement detection. The same phenomenon of a
decreasing noise tolerance when including more correlations was observed.

The preceding observations concerning the comparison of the measures
$\mathcal{F}$ and $\mathcal{D}$ can be understood by noting that the relative
entropy uses all information contained in the probability distributions for the
measurement outcomes, whereas the quantity $\mathcal{F}$ effectively reduces
each probability distribution to one parameter.

We recall that the value of the relative entropy $\mathcal{D}$ has an
interpretation in terms of probabilities (Section~\ref{sec:relent}). This is in
contrast to the quantity $\mathcal{F}$, whose numerical value is fixed only by
a normalization condition on the observables (Section~\ref{sec:fidelity}).
While in certain cases it can be interpreted as a noise tolerance and it is
useful for comparing different observables, it gives no quantitative statement
about the discrepancy between the prepared state and states which we want to
exclude. Finally, we note that in experimental applications also the error
estimates for either quantity must be taken into account. Though we have done
so in Eqs.~\eqref{eq:exp1}--\eqref{eq:exp4}, a systematic analysis of this
point is beyond the scope of our present article.

\section{Discriminating three-qubit states}
\label{sec:three}

Now we consider the three-qubit case, aiming at the discrimination of the
three-qubit GHZ state and the three-qubit W state. These two states are
relevant as representatives of the two different entanglement classes of
genuine three-qubit entanglement~\cite{DuVC00}.

The three-qubit GHZ state is given by
\begin{equation}
  \ket{\text{GHZ}_3}=\frac{1}{\sqrt{2}}\bigl(\ket{000}+\ket{111}\bigr).
\end{equation}
As in the four-qubit case, this state can be described by its stabilizing
operators. The stabilizer group is given by the eight observables
\begin{multline}
  S_{\text{GHZ}_3}=\{\one\one\one,\ \one ZZ\text{ and perm.},\\
  XXX,\ -XYY\text{ and perm.}\}.
\end{multline}
The three-qubit W state
\begin{equation}
  \ket{\text{W}_3}=\frac{1}{\sqrt{3}}\bigl(\ket{001}+\ket{010}+\ket{100}\bigr)
\end{equation}
is not a stabilizer state. If we expand its density matrix into Pauli matrices
we arrive at
\begin{equation}
  \label{eq:wcorr}
  \begin{split}
    &\ketbra{\text{W}_3}
    =\frac{1}{24}\bigl[3\cdot\one\one\one+(\one\one Z+\text{perm.})\\
    &{+2}(\one XX+\text{perm.})+2(\one YY+\text{perm.})-(\one ZZ+\text{perm.})\\
    &{+2}(XXZ+\text{perm.})+2(YYZ+\text{perm.})-3\cdot ZZZ\bigr].
  \end{split}
\end{equation}

\subsection{Discriminating the GHZ state from the W state}

Again, we will first compute $\mathcal{F}$ and $\mathcal{D}$ for the case that
all stabilizing operators (except for $\one$) of the GHZ state are used, and
afterwards look for optimal families of observables.

Parameterizing local unitaries as $U(\varphi,\theta,\psi)
=\exp(\ir\psi\sigma_z/2)\exp(\ir\theta\sigma_y/2)\exp(\ir\varphi\sigma_z/2)$,
we obtain $\bra{\text{W}_3}U^\dagger\,\one ZZ\, U\ket{\text{W}_3}
=\frac{1}{3}[-\cos(\theta_2)\cos(\theta_3)
+2\cos(\varphi_2-\varphi_3)\sin(\theta_2)\sin(\theta_3)]$. The expectation values
of $Z\one Z$ and $ZZ\one$ can be obtained by cyclically permuting the indices
of the angles, as the W state is permutationally invariant. We find
$\max_\text{LU}\bra{\text{W}_3}\one ZZ\ket{\text{W}_3}=2/3$, where the maximum
is attained when $\cos(\varphi_2-\varphi_3)\sin(\theta_2)\sin(\theta_3)=1$. Thus
the expectation values of $\one ZZ$, $Z\one Z$, and $ZZ\one$ can be maximized
simultaneously. We choose the solution $\varphi_1=\varphi_2=\varphi_3=0$ and
$\theta_1=\theta_2=\theta_3=\pi/2$.

Let us now consider the remaining stabilizing operators. Assuming the above
choice for the angles $\varphi_i$ and $\theta_i$ and making the symmetry
assumption $\psi_1=\psi_2=\psi_3=\psi$, we have
$\bra{\text{W}_3}U^\dagger\,XXX\,U\ket{\text{W}_3}=3\cos^3(\psi)-2\cos(\psi)$ and
$\bra{\text{W}_3}U^\dagger(-XYY)U\ket{\text{W}_3}
=3\cos^{3}(\psi)-\frac{7}{3}\cos(\psi)$ and finally
\begin{multline}
  D_{\substack{XXX,-XYY,\\-YXY,-YYX}}
  \bigl(\text{GHZ}_3\big\Vert U\ket{\text{W}_3}\bigr)\\
  =-\log\Bigl[\frac{1}{8}\bigl(1+3z^{3}-2z\bigr)
  \bigl(1+3z^{3}-\frac{7}{3}z\bigr)^{3}\Bigr],
\end{multline}
where $z=\cos(\psi)$. This function is minimal at $z=-1/2$. 

In conclusion, we found that
\begin{align}
  \mathcal{D}_{S^*_{\text{GHZ}_3}}(\text{GHZ}_3\Vert\text{W}_3)
  &=\frac{1}{7}\bigl(-3\log\frac{5}{6}-\log\frac{13}{16}
  -3\log\frac{43}{48}\bigr)\nonumber\\
  &\approx 0.2235.
  \label{eq:ghzwd}
\end{align}
While our analytical calculation required some symmetry assumptions, this value
is also obtained by numerical minimization over all Euler angles.

For the fidelity-based measure $\mathcal{F}$ we note that the rotation we found
when minimizing $\mathcal{D}$ gives
$\abs{\bra{\text{GHZ}_3}U\ket{\text{W}_3}}^2=3/4$. This is known to be the
highest possible overlap of the GHZ state with local unitaries (or, indeed,
SLOCC) of the W state~\cite{ABLS01}. With Eq.~\eqref{eq:fid} we obtain
\begin{equation}
  \label{eq:ghzwf}
  \mathcal{F}_{S^*_{\text{GHZ}_3}}(\text{GHZ}_3\Vert\text{W}_3)
  =\frac{2}{7}.
\end{equation}
In Ref.~\cite{ABLS01} the state
\begin{equation}
  \ket{\widehat{\text{W}}_3}=\frac{1}{2\sqrt{6}}(3,-1,-1,-1,-1,-1,-1,3)
\end{equation}
was found to maximize the overlap of $\ket{\text{GHZ}_3}$ with the SLOCC class
of the W state. This state is in fact LU equivalent to $\ket{\text{W}_3}$.
Though $\ket{\widehat{\text{W}}_3}$ minimizes $F$, it does not minimize $D$, as
it gives the value $D_{S^*_{\text{GHZ}_3}}(\text{GHZ}_3\Vert\widehat{\text{W}}_3)
=-\frac{6}{7}\log\frac{5}{6}\approx 0.2255$.

We want to find families of observables that give the highest values of
$\mathcal{F}$ and $\mathcal{D}$. We claim that any combination of
\begin{equation}
  \one ZZ,\ Z\one Z,\ ZZ\one 
\end{equation}
is optimal in this sense among all GHZ stabilizing operators. As we have seen
above, any element of this family gives $\mathcal{F}=1/3$ and
$\mathcal{D}=-\log(5/6)$. We only have to show that for any other stabilizing
operator, or combination of stabilizing operators, there exists a local unitary
such that $F\le 1/3$ and $D\le -\log(5/6)$. Taking in the above calculations
$\cos(\psi)=1$ we have $\bra{\text{W}_3}U^\dagger\,XXX\,U\ket{\text{W}_3}=1$
and $\bra{\text{W}_3}U^\dagger(-XYY)U\ket{\text{W}_3}=2/3$. This proves our
claim.

One might ask if these optimal families of observables can be used for the
construction of a witness operator for the GHZ entanglement class. This is,
however, not the case, as
\begin{equation}
  \max_\text{SLOCC of $\text{W}_3$}
  \bra{\text{W}_3}(\one ZZ+Z\one Z+ZZ\one)\ket{\text{W}_3}
  =3
\end{equation}
holds, which can be seen from the fact that the fully separable state
$\ket{000}$ gives this value.

We summarize the main results of this subsection:

\begin{observation}
  If all GHZ stabilizing operators except the identity are used for the
  discrimination of the GHZ state from all LU equivalents of the W state, the
  measures $\mathcal{D}$ and $\mathcal{F}$ are given by Eqs.~\eqref{eq:ghzwd}
  and~\eqref{eq:ghzwf}. If we consider subsets of the stabilizer group, any
  combination of $\one ZZ$, $Z\one Z$, and $ZZ\one$ is an optimal family of
  observables, yielding $\mathcal{F}=1/3$ and $\mathcal{D}=-\log(5/6)$.
\end{observation} 

\subsection{Discriminating the W state from the GHZ state}

For the reverse problem, the discrimination of $\ket{\text{W}_3}$ from the
local unitaries of $\ket{\text{GHZ}_3}$, there is no obvious choice of
observables, as the W state is not a stabilizer state. However, each of the
observables
\begin{equation}
  \one\one Z,\ \one Z\one,\ Z\one\one
\end{equation}
has an expectation value of $1/3$ for the W state and zero expectation value
for all local unitaries of $\ket{\text{GHZ}_3}$, as for the GHZ state all
reduced one-qubit density matrices are maximally mixed. Thus any appropriately 
normalized combination of
the above observables gives $\mathcal{F}=1$, which is the optimal value. 

All of these combinations give
\begin{equation}
  \label{eq:value}
  \mathcal{D}(\text{W}_3\Vert\text{GHZ}_3)
  =\frac{2}{3}\log\frac{4}{3}-\frac{1}{3}\log\frac{3}{2}
  \approx 0.0817.
\end{equation}
This is the best possible value among all operators occurring in the
decomposition of $\ket{\text{W}_3}$ in Eq.~\eqref{eq:wcorr}. To see this, we
choose the rotation $\varphi=2\pi/3$, $\theta=3\pi/2$, and $\psi=5\pi/4$ on all
qubits and observe that all these operators give a value of $D$ less or equal
to the one in Eq.~\eqref{eq:value}.

\section{General graph states}
\label{sec:graph}

In the previous sections we observed that the stabilizing operators of a given
state are natural candidates for observables that discriminate this state from
other states. The graph states are a large class of states which are
unambiguously described by their stabilizing operators~\cite{HDER06}. In this
section we will derive some general statements about the discrimination of
graph states.

\begin{figure}
  \centering
  \includegraphics[width=0.45\textwidth]{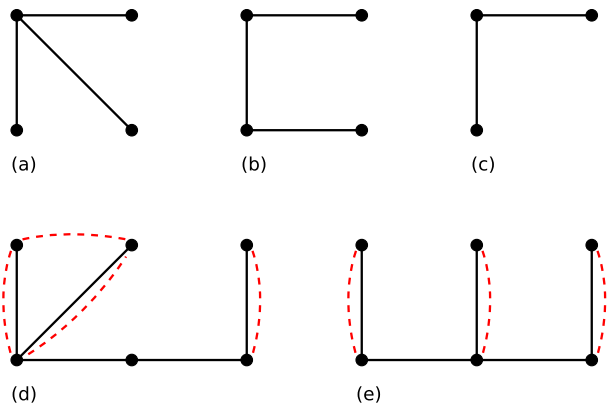}
  \caption{\label{fig:graphs} Graphs of the graph states appearing in the text.
  (a)~Four-qubit GHZ state, (b)~four-qubit linear cluster state,
  (c)~three-qubit GHZ state, (d) and (e)~the graph states discussed at the end
  of Section~\ref{sec:graph}. In (d) and (e) dashed red lines denote the
  presence of two-point correlations.}
\end{figure}

A graph state is described by its graph, that is, a collection of vertices,
each standing for a qubit, and edges connecting vertices. Some graphs are
depicted in Fig.~\ref{fig:graphs}. With each vertex (or qubit) $i$ we associate
the correlation operator $K_i$ defined as the observable that acts as $X$ on
the vertex $i$ and as $Z$ on every vertex in the neighborhood $N(i)$ of $i$,
i.\,e., all vertices connected to $i$ by an edge
\begin{equation}
  K_i=X_i\prod_{j\in N(i)}Z_j.
\end{equation}
The graph state $\ket{G}$ is defined as the unique common eigenstate of all
$K_i$ with eigenvalue $+1$. The products of the $K_i$ form a commutative group
$S$, and the graph state is an eigenstate of all elements of this group. The
group $S$ is called the stabilizer group of $\ket{G}$ and its elements the
stabilizing operators. For an $n$-qubit state, that stabilizer group has $2^n$
elements. The sum of all stabilizing operators $S_i\in S$ gives the projector
onto the graph state
\begin{equation}
  \ket{G}\bra{G}=\frac{1}{2^n}\sum_{i=1}^{2^n}S_i,
\end{equation}
and Eq.~(\ref{eq:gsproj}) is an example of this general property. The
stabilizing operators thus contain a description of all correlations present in
the state.

Alternatively from describing perfect correlations, the graph can be understood
as a description of the interactions that created the state~\cite{HDER06}:
Every edge corresponds to an Ising-type interaction and the state is obtained
by applying these interactions to the product state
$\ket{+}\ket{+}\cdots\ket{+}$. In addition, it should be noted that it is
possible for two different graphs to lead to states which are equivalent under
local unitary transformations and hence have the same entanglement
properties~\cite{HDER06}.

As an example, we consider the graph in Fig.~\ref{fig:graphs}~(a). The
generating operators of the stabilizer group are
\begin{equation}
  K_1=XZZZ,\ K_2=ZX \one\one ,\ K_3=Z\one X\one,\ K_4=Z\one\one X.
\end{equation}
After a local change of bases we recognize the graph state as the four-qubit
GHZ state with the complete stabilizer group given in Eq.~\eqref{eq:ghzstab}.
The four-qubit linear cluster state and the three-qubit GHZ state are given by
the graphs in Figs.~\ref{fig:graphs}~(b) and \ref{fig:graphs}~(c).

In our discussion on the discrimination of the GHZ from the cluster state we
learned that the optimal families of observables consist of two-point
stabilizing operators. The reason for their optimality is that the cluster
state has fewer two-point stabilizing operators than the GHZ state. Hence one
may try do derive general results depending only on the numbers of two-point
(or higher-order) stabilizing operators.

The number of two-point correlations of a graph state can easily be obtained
from its graph~\cite{HyGS09, GHG10}. Restricting ourselves to connected graphs
with three or more vertices, there are three possibilities to obtain two-point
stabilizing operators:
\begin{enumerate}
  \item Vertices connected to the rest of the graph by only one edge. The
    generator associated to such a vertex is a two-point operator of the form
    $XZ$.
  \item\label{it:xx} Pairs of unconnected vertices whose neighborhoods are
    equal: $N(i)=N(j)$. The product of the two generators associated to such a
    pair has the form $XX$.
  \item\label{it:yy} Pairs of connected vertices for which
    $N(i)\cup\{i\}=N(j)\cup\{j\}$. This means that their neighborhoods apart
    from $i$ and $j$ are the same. The product of their generators has the form
    $YY$.
\end{enumerate}
The product of the generators associated to vertices $i$ and $j$ is never equal
to the identity at positions $i$ and $j$. Therefore it is a two-point
stabilizing operator if and only if it is equal to the identity at all other
positions, which leaves only the possibilities~\ref{it:xx} and~\ref{it:yy}. For
the same reason the product of three or more generators is never a two-point
stabilizing operator. This shows that the above list exhausts all possibilities
to obtain a two-point stabilizing operator.

Now, let $\ket{G_1}$ and $\ket{G_2}$ be two graph states, $k_1$ and $k_2$ the
numbers of two-point correlations of these states, and let $P_{G_1}$ be the set
of two-point stabilizing operators of $\ket{G_1}$. We assume $k_1>k_2$ (our
result will be trivial otherwise). We can then derive a lower bound on
$\mathcal{F}_{P_{G_1}}(G_1\Vert G_2)$ and $\mathcal{D}_{P_{G_1}}(G_1\Vert G_2)$
that depends only on the numbers $k_1$ and $k_2$. Namely, we have
\begin{align}
  \mathcal{F}_{P_{G_1}}(G_1\Vert G_2)&\ge\frac{k_1-k_2}{k_1},\\
  \mathcal{D}_{P_{G_1}}(G_1\Vert G_2)&\ge\frac{k_1-k_2}{k_1}.
\end{align}
To see this, note that from the above discussion it is clear that all two-point
stabilizing operators of the same graph state act on different pairs of
qubits~\cite{GHG10}. It follows that any two-point stabilizing operator of
$\ket{G_2}$ can have nonzero overlap with at most one stabilizing operator of
$\ket{G_1}$. This shows that at least $k_1-k_2$ two-point stabilizing operators
of $\ket{G_1}$ will have zero overlap with $\ket{G_2}$, and thus give
$\mathcal{F}=\mathcal{D}=1$. Note that this still holds if local unitaries and
permutations of qubits are considered.

For $\ket{G_1}=\ket{\text{GHZ}_4}$ and $\ket{G_2}=\ket{\text{C}_4}$ the bounds
give the exact results. This is, however,  not always the case, as the example
of Figs.~\ref{fig:graphs}~(d) and \ref{fig:graphs}~(e) shows: Here, $k_1=4$ and $k_2=3$. Of the
two-point correlations of $\ket{G_1}$, three connect three qubits in a
triangle, while for $\ket{G_2}$ the connected pairs are all disjoint. This
shows that at most two of the two-point stabilizing operators of $\ket{G_1}$
can have nonzero overlap with $\ket{G_2}$. In this example,
$\mathcal{F}_{P_{G_1}}(G_1\Vert G_2) =\mathcal{D}_{P_{G_1}}(G_1\Vert
G_2)=1/2>1/4=(k_1-k_2)/k_1$.

While one can also use higher-order stabilizing operators for the
discrimination, it is more difficult to derive general results for them.
Nevertheless, for two given graph states the number of three-point (or
higher-order) correlations can directly be computed by writing down the whole
stabilizer group; one may then compare the different numbers of higher order
stabilizing operators.

\section{Discussion and conclusion}
\label{sec:discussion}

In this article, we have developed strategies for showing that an
experimentally prepared state is not in a certain class of undesired states,
given by all local unitaries, or all local unitaries and permutations of qubits,
of another state. 

We introduced two measures for the discrimination strength of observables. The
first measure, denoted by $\mathcal{F}$, was based on the difference of the
expectation values of the observable on either state. It could be interpreted
as a noise tolerance, similar to the case of witness operators. The other
measure, denoted by $\mathcal{D}$, was defined as the relative entropy of the
probability distributions for the measurement outcomes on either state. It
gives a quantitative answer to the discrimination problem by allowing to
compare the likelihood that the experimental data origin from the undesired
state with, e.\,g., the likelihood that a sequence of highly biased coin tosses can
be explained by a fair coin. In particular, the measure $\mathcal{D}$ is
defined in such a way that a maximization of it corresponds to maximizing the
discrimination strength while keeping the total number of measurement runs
constant, thus answering the question for the best discrimination with limited
experimental resources.

We discussed in detail the discrimination of the four-qubit GHZ state from the
four-qubit linear cluster state, and vice versa, using stabilizer observables.
We also discussed the three-qubit GHZ state and the W state, thus showing that
our method can be generalized to states that are no graph states. Finally, we
derived a general result on the discrimination of two graph states with the
two-point stabilizing operators of one of them, using the power of the graph
formalism.

For a specific example we studied the noise tolerance of the quantities
$\mathcal{F}$ and $\mathcal{D}$ and their performance for experimental data.
Concerning the comparison of the two measures, our observation that the measure
$\mathcal{D}$ is more robust against noise than $\mathcal{F}$ could be
explained by the fact that the relative entropy uses all information contained
in the probability distributions, while the measure $\mathcal{F}$ effectively
reduces each probability distribution to one parameter (namely, the sum of the
expectation values). Thus, our conclusion is as follows: Either measure can be
used to compare the suitability of different observables for a given
discrimination task. For the evaluation of experimental results the relative
entropy $\mathcal{D}$ is to be preferred, since it uses all available
information and allows for a clear statistical interpretation.

There are several interesting open questions and possible generalizations for
the future: First, one could extend our analysis to the measurement of
nondichotomic observables or arbitrary product bases [see our discussion of
Eq.~\eqref{eq:bases}]. Second, one could consider SLOCC equivalence classes
instead of LU classes. Our measures $\mathcal{F}$ and $\mathcal{D}$ are equally
applicable for that case, only the optimization is different. Third, one could
connect our results to other methods for characterizing multipartite
entanglement classes. For instance, there exist witnesses distinguishing the
class of mixed three-qubit GHZ states from the class of mixed W
states~\cite{ABLS01}. Here, mixed W states are those three-qubit states that can
be written as a mixture of pure states that are biseparable or SLOCC equivalent
to the W state. The discrimination of such classes of mixed states is a
different problem than the one considered in this article; nevertheless, it would
be interesting to understand possible connections. 

\begin{acknowledgments}
  We thank O.~Gittsovich, B.~Jungnitsch, B.~Kraus, E.~Solano, and G.~T\'oth for
  discussions. This work has been supported by the FWF (START Prize and SFB
  FOQUS) and the EU (NAMEQUAM, QICS).
\end{acknowledgments}

\appendix

\section{Properties of the relative entropy}
\label{app:relent}

The relative entropy, also called Kullback-Leibler divergence, from the
probability distribution $P=\{p_1,\ldots,p_m\}$ to the probability distribution
$Q=\{q_1,\ldots,q_m\}$ is defined as~\cite{CoTh91}
\begin{equation}
  D(P\Vert Q)=\sum_{i=1}^mp_i\log\bigl(\frac{p_i}{q_i}\bigr).
\end{equation}
We use the logarithm to the base of two, $\log=\log_2$, and define $0\log(0)=0$
(more explicitly: $0\log(\frac{0}{q})=0$, $p\log(\frac{p}{0})=\infty$, and
$0\log(\frac{0}{0})=0$). We list without proof the properties of the relative
entropy that are needed in this article:
\begin{enumerate}
  \item Positive definiteness: $D(P\Vert Q)\ge 0$ with equality if and only if
    $P=Q$.
  \item $D(P\Vert Q)=\infty$ if and only if $q_i=0$, but $p_i>0$ for some $i$,
    that is, if an event is impossible according to the probability
    distribution $Q$, but occurs with a nonvanishing probability according to
    $P$.
  \item $D(P\Vert Q)\neq D(Q\Vert P)$ in general, so $D$ is not a metric.
  \item\label{it:grouping} Grouping rule: If we decide to consider events 1 and
    2 as one event, leaving us with probability distributions
    $\tilde{P}=\{p_1+p_2,p_3,\ldots,p_m\}$ and
    $\tilde{Q}=\{q_1+q_2,q_3,\ldots,q_m\}$, then
    \begin{equation}
      D(P\Vert Q)=D(\tilde{P}\Vert\tilde{Q})+(p_1+p_2)D(P'\Vert Q'),
    \end{equation}
    where $P'=\{p_1/(p_1+p_2),p_2/(p_1+p_2)\}$ and
    $Q'=\{q_1/(q_1+q_2),q_2/(q_1+q_2)\}$. Note that $D(P'\Vert Q')$ is the
    relative entropy within the ``box'' containing events 1 and 2, and
    $(p_1+p_2)$ is the $P$ probability of obtaining an event from this box.
  \item\label{it:additivity} If $P$ and $Q$ are the joint probability
    distributions for two independent random variables,
    $p_{ij}=p^{(1)}_ip^{(2)}_j$ and $q_{ij}=q^{(1)}_iq^{(2)}_j$, then
    \begin{equation}
      D(P\Vert Q)=D(P^{(1)}\Vert Q^{(1)})+D(P^{(2)}\Vert Q^{(2)}).
    \end{equation}
\end{enumerate}

\bibliography{strategies}

\end{document}